\documentclass[12pt,english]{elsart}
\usepackage[T1]{fontenc}
\usepackage[latin1]{inputenc}
\usepackage{amsmath}
\usepackage{graphicx}
\usepackage{subfigure}
\usepackage{babel}
\makeatother
\begin{document}
\begin{frontmatter}
\title{The influence of asymmetry on a magnetized proto-neutron star.}
\author{I. Bednarek, A. Brzezina, R. Ma\'{n}ka, M. Zastawny-Kubica}
\address[Katowice]
{ Department of Astrophysics and Cosmology, Institute of Physics, \\
  University of Silesia, Uniwersytecka 4, PL-40-007 Katowice, Poland.}

\maketitle
\begin{abstract}
Using the Relativistic Mean Field Theory (RMF) it is shown that
different proton fraction which is directly connected with the
neutron excess and with the asymmetry of the system affects
proto-neutron stars parameters and changes their composition. The
obtained form of the equation of state allows to construct the
mass-radius relations and shows that the increasing asymmetry
creates more compact stars. The inclusion of $\delta $ meson
together with nonlinear vector meson interaction terms and
magnetic field make this effect even stronger.

\end{abstract}
\end{frontmatter}
\section*{Introduction}
Properties of dense matter in strong magnetic field has been the
subject of investigations in astrophysics of white dwarfs,
proto-neutron and neutron stars. Observations of pulsars suggest
large surface fields of the order of $10^{14}$ $G$  \cite{web}.
This very high value indicates the existence of even stronger
interior magnetic fields. In fact the virial theorem suggests that
interior magnetic fields can reach the value of the order of
$10^{18}$ $G$. The influence of this extraordinary high magnetic
field on neutron star matter properties are of particular
importance after the discovery of magnetars  \cite{web}. Their
observations suggest that in the case of these objects we are
dealing with young neutron stars with extremely strong surface
magnetic fields $\sim$ $10^{15}$ $G$ which in turn give an
interior field of the order of $10^{18}$ $G$. Also properties of
proto-neutron stars, under the influence of magnetic field should
be examined. Assumptions concerning composition and the equation
of state of hot, lepton rich matter have been studied by many
authors (Strobel et al. 1999, Bombaci et al. 1995, Takatsuka 1995,
Ma\'{n}ka et al. 2001) \cite{Strob,Bomb,Takat,mil4}. Matter inside
a proto-neutron star is highly degenerate and chemical potentials
of its constituents are a few hundreds of $MeV$. Like in a neutron
star the strength of magnetic field of a proto-neutron star
changes from $10^{8}$ $G$ at the surface up to $10^{18}$ $G$ in
the center \cite{pal}.
 A proto-neutron star \cite{pra1,pra2,pra3} is a result of
a supernova explosion which forms low entropy core with trapped
neutrinos. The core is surrounded by a low density, high entropy
mantle. The construction of a proto-neutron star model is based on
various realistic equations of state and gives a general picture
of proto-neutron star interiors. The more complete and realistic
description of a proto-neutron star requires taking into
consideration effects of finite temperature and nonzero magnetic
fields. Using the relativistic mean-field theory approach the
adequate form of the equation of state enlarged by contributions
coming from magnetic field and temperature is constructed and
serves as an input to the Oppenheimer-Volkoff-Tolman equations.
The theory considered here comprises electrons, neutrinos, scalar,
vector-scalar and vector-isovector mesons. The RMF theory implies
that the nucleon interactions appear through the exchange of meson
fields \cite{ser,rei,toki,bedn}. This theory is very useful in
describing properties of nuclear matter and finite nuclei. Its
extrapolation to large charge asymmetry is of considerable
interest in nuclear astrophysics and particulary in constructing
proto-neutron and neutron star models where extreme conditions of
isospin are realized. The model considered describes high isospin
asymmetric matter and it has to be extended by the inclusion of
isovector-scalar meson $a_{0}\,(980)$ (the $\delta $ meson) and
nonlinear vector meson interaction terms. Knowing the form of the
equation of state is the decisive factor in determining properties
of proto-neutron stars such as: central density, mass-radius
relation, crust extent or the moment of inertia. The essential
goal of this paper is to obtain the equation of state for
proto-neutron star matter within the described above model in the
relativistic mean field approach (RMF) and to study the influence
of proton fraction on proto-neutron stars parameters. The variable
proton fraction together with the inclusion of $\delta $ meson and
nonlinear vector meson interactions alter the proto-neutron stars
chemical composition. This in turn affects the properties of the
star making the proto-neutron star more compact.
\section*{The generalized Relativistic Mean Field approach}
The Lagrangian function for the system can be written as a sum of
a baryonic part $\mathcal{L}_{B}$ which includes baryon-meson
interaction terms, mesonic part $\mathcal{L}_{\mathcal{M}}$
containing additional interactions between mesons which
mathematically express themselves as supplementary, nonlinear
terms in the Lagrangian function, the leptonic part $\mathcal{L}_{L}$,
the Lagrangian density function of the QED theory $\mathcal{L}_{QED}$
and the gravitational term $\mathcal{L}_{G}$
\begin{equation}
\mathcal{L}=\mathcal{L}_{B}+\mathcal{L}_{L}+\mathcal{L}_{\mathcal{M}}+\mathcal{L}_{G}+\mathcal{L}_{QED}.\label{lagra}
\end{equation}
The final form of the Lagrangian function is given by
\begin{eqnarray}
\label{lagrangian}
 & \mathcal{L}=i\overline{\psi }\gamma ^{\mu }D_{\mu }\psi
-\overline{\psi }(M_N-g_{\sigma }\varphi -I_{3N}g_{\delta }\tau
^{a}\delta ^{a})\psi -\kappa _{N}\overline{\psi }\sigma _{\mu \nu
}F^{\mu \nu }\psi & \\ \nonumber
 & - \frac{1}{4}c_{3}(\omega _{\mu }\omega ^{\mu
})^{2} - \frac{1}{4}R_{\mu \nu }^{a}R^{a\mu \nu
}-\frac{1}{2}M_{\rho }^{2}\rho _{\mu }^{a}\rho ^{a\mu
}+\frac{1}{2}\partial _{\mu }\delta ^{a}\partial ^{\mu }\delta
^{a}-\frac{1}{2}M_{\delta }^{2}\delta ^{a}\delta ^{a} & \\
\nonumber & + i\sum _{f=1}^{2}\bar{L}_{f}\gamma ^{\mu
}\widetilde{D}_{\mu }L_{f}-\sum
_{f=1}^{2}g_{f}(\bar{L}_{f}He_{Rf}+h.c)-
 \frac{1}{2}\partial _{\mu }\varphi \partial ^{\mu }\varphi
-U(\varphi ) & \\ \nonumber
 & - \frac{1}{4}W_{\mu \nu }W^{\mu \nu
}-
 \frac{1}{2}
M_{\omega }^{2}\omega _{\mu }\omega ^{\mu } - \frac{1}{4}F_{\mu
\nu }F^{\mu \nu }+ \frac{1}{4!}g_{\rho}^4\zeta (b_{\mu
}^{a}b^{a\mu })^2 & \\ \nonumber &
+(g_{\rho}g_{\omega})^2\Lambda_vb_{\mu}^ab^{a\mu}\omega_{\mu}\omega^{\mu}+(g_{\rho}g_{s})^2\Lambda_4b_{\mu}^ab^{a\mu}\varphi^2
& \\ \nonumber
\end{eqnarray}
with the covariant derivatives $D_{\mu }$ and  $\widetilde{D}_{\mu
}$ defined as
\begin{eqnarray}
D_{\mu }=\partial _{\mu }+\frac{1}{2}ig_{\rho }\rho _{\mu
}^{a}\sigma ^{a}+ig_{\omega }\omega _{\mu }+iQA_{\mu
},\hspace{0.7cm} \widetilde{D}_{\mu }=\partial _{\mu }+iQA_{\mu }.
\end{eqnarray}
 In the case of nucleons $Q$ takes the value $e$ for protons ($e$
is the electron charge) and $0$ for neutrons ($Q={e,0}$). In the
leptonic sector ($Q={-e,0}$) where $-e$ is given for electrons and
muons and $0$ for neutrinos. $H$ is Higgs field and this field has
only the residual form
\begin{eqnarray*}
H=\frac{1}{\sqrt{2}}\left(\begin{array}{c}
 0\\
 v\end{array}
\right)
\end{eqnarray*}
where the value $v=250 \,MeV$ comes from the electroweak
interaction scale. In a strong magnetic field, contributions
coming from the anomalous magnetic moments of protons and neutrons
($\kappa _{N},\,\,N=\{p,n\}$) have to be also considered. The
anomalous magnetic moments introduced via the minimal coupling of
nucleons to the electromagnetic field tensor can be represented
phenomenologically by interactions of the type $\kappa
_{N}\overline{\psi }\sigma _{\mu \nu }F^{\mu \nu }\psi $ , where
$\sigma _{\mu \nu }=\frac{i}{2}\{\gamma _{\mu },\gamma _{\nu }\}$
\cite{Lati}. Considering both the physical conditions and chemical
equilibrium which are likely to be satisfied in a proto-neutron
star one can construct a proto-neutron star model with a
temperature $T$ equals $20$ $MeV$ and magnetic field strength
$B\sim 10^{3}B_{c}^{e}$ where $B_{c}^{e}$ is the value of the
critical magnetic field for an electron $B_{c}^{e}$=
$m_{e}^{2}/\mid e\mid $= $4.414\times 10^{13}$ $G$. The model
examined in this paper represents the one with moderate value of
magnetic field. In this case the contributions to the equation of
state coming from anomalous magnetic moments are small $(\sim
1MeV)$ in comparison with the energy of magnetic field. Thus this
relatively small value of magnetic field allows to neglect the
anomalous magnetic moments unlike to the case describe in the
paper by Prakash et al. \cite{Lati} where authors consider the
case of extremely strong magnetic field of the order of $10^{18}$
$G$. Realistic proto-neutron star models describe electrically
neutral, hot, high density matter being in $\beta $ equilibrium.
The last condition implies the presence of leptons which is
expressed by adding the Lagrangian of leptons to the Lagrangian
function. The fermion fields which are included in this model
consists of neutrons, protons, electrons, muons and neutrinos
\begin{equation}
\psi =\left(\begin{array}{c}
 \psi _{p}\\
 \psi _{n}\end{array}
\right),\, \, L_{1}=\left(\begin{array}{c}
 \nu _{e}\\
 e^{-}\end{array}
\right)_{L},\, \, L_{2}=\left(\begin{array}{c}
 \nu _{\mu }\\
 \mu ^{-}\end{array}
\right)_{L},\, \, e_{Rf}=\left(\begin{array}{cc}
 e_{R}^{-},\, \mu _{R}^{-} \end{array}
\right).
\end{equation}
As a proto-neutron star matter is of sizeable asymmetry the
additional $\delta$ meson has been included and the meson sector
is composed of isoscalar (scalar $\sigma $ and vector $\omega $)
and isovector (scalar $\delta $ and vector $\rho $) mesons. The
potential function $U(\varphi )=\frac{1}{2}M_{\sigma }^{2}\varphi
^{2}+\frac{1}{3}g_{2}\varphi ^{3}+\frac{1}{4}g_{3}\varphi ^{4}$
has a very well known form introduced by Boguta and Bodmer
\cite{bod}.
\begin{table}
\centerline{Table 1a} \vspace*{0.15in} \centerline{The parameter
sets of the model \cite{Lati,osa7}.} \vspace*{0.15in}
\begin{center}
\begin{tabular}{|c|c|c|c|c|c|}
\hline
 $ $&
 $g_{\rho }$&
 $g_{\delta }$&
 $g_{\sigma }$&
 $g_{\omega}$ &
 $M_{\sigma}\,\,(MeV)$\\
 \hline $TM1$&
 $9.264$&
 $0$&
 $10.0289$&
 $12.6139$ &
 $511.12$\\
 \hline $TM1+nonl.$&
 $10.875$&
 $3.5$&
 $10.0289$&
 $12.6139$ &
 $511.12$\\
\hline $GM3$ & $8.5417$ & $0$ & $7.1857$ & $8.7041$ & $450.00$\\
\hline
\end{tabular}
\end{center}
\vspace*{0.15in}
\end{table}
\begin{table}
\centerline{Table 1b} \vspace*{0.15in} \centerline{The
self-interacting coupling constants \cite{Lati,osa7}.}
\vspace*{0.15in}
\begin{center}
\begin{tabular}{|c|c|c|c|c|c|c|}
\hline $ $ &
 $g_{2}$\, (MeV)&
 $g_{3}$&
 $c_{3}$&
 $\Lambda_{v}$&
 $\Lambda_{4}$&
 $\zeta$\\
\hline $TM1$& $1427.18$ &
 $0.6183$&
 $71.3075$&
 $-$&
 $-$&
 $-$ \\
\hline $TM1+nonl.$& $1427.18$&
 $0.6183$&
 $71.3075$&
 $0.008$&
 $0.001$&
 $0.5$ \\
\hline $GM3$& $3016.52$& $-6.4546$ & $0$ & $-$ & $-$ & $-$ \\
\hline
\end{tabular}
\end{center}
\vspace*{0.15in}
\end{table}
Nucleon masses are denoted by $M_{N}$ (N=p,n) whereas
$M_{\omega}$, $M_{\rho}$, $M_{\sigma}$ and $M_{\delta}$ are masses
assigned to the meson fields. They are taken at their
experimentally values: $M_{\omega}=783\,MeV$, $M_{\rho}=770\,MeV$
and $M_{\delta}=980\, MeV$. The nonlinear vector meson interaction
terms have been added in order to give a detailed account of a
matter inside a proto-neutron star. This has been done through the
construction of adequate form of the equation of state. To
determine the equation of state at finite temperature and
nonvanishing magnetic field some parameterizations have been used.
The first one is known as $TM1$ parameter set \cite{osa7}, the
second one  noted as $TM1+ nonl.$ includes nonlinear terms in the
meson sector. The third is the $GM3$ parameter set \cite{Lati}.
The parameters entering the Lagrangian function are the coupling
constants $g_{\rho }$, $g_{\delta }$, $g_{\sigma }$, $g_{\omega}$
for meson fields and self-interacting coupling constants $g_{2}$
$g_{3}$, $c_{3}$, $\Lambda_{v}$, $\Lambda_{4}$ and $\zeta$. All
these parameters have been chosen to reproduce properties of
symmetric nuclear matter at saturation and they are collected in
Tables 1a and 1b. In the original $GM3$ and $TM1$ approaches
neutrinos are not included since these parameterizations describe
a neutron star matter. In this case neutrinos leave the star
unhindered. In order to construct the neutron star model through
the entire density span the addition of the equations of state,
characteristic for the inner and outer core, relevant for lower
densities is necessary.  In the result the composite equation of
state for which the TM1 or GM3 parameter groups  describing the
neutron star core are supplemented with the Negele-Voutherin (NV)
+ Bonn describing inner crust can be used. The equilibrium
properties of nuclear matter which are known with reasonable
precision are: the nuclear saturation density $n_{S}$, the binding
energy $E_{b}$, the incompressibility $K_{0}$, the symmetry energy
$a_{sym}$ and the effective nucleon mass. Nuclear matter
properties calculated for given parameter sets (Tables 1a and 1b)
are summarized in Table 2. The presence of additional nonlinear
$\rho$ meson interaction terms and $\delta$ meson which carries
izospin contribute to the symmetry energy. Thus the value of the
parameter $g_{\rho}$  has to be redefined for the $TM1+nonl.$
parameterization. Parameters $g_{\rho}$, $g_{\delta}$,
$\Lambda_{v}$, $\Lambda_{4}$ and $\zeta$ are not chosen arbitrary
they have to be given in a specified form adequate to reproduce
the value of the symmetry energy $a_{sym}$.
\begin{table}
\centerline{Table 2} \vspace*{0.15in} \centerline{The nuclear
matter properties.} \vspace*{0.15in}
\begin{center}
\begin{tabular}{|c|c|c|c|c|c|}
\hline $ $&
 $E_{b}\, (MeV)$&
 $m_{eff}/M$&
 $n_{S}\, (fm^{-3})$&
 $K_{0}\, (MeV)$&
 $a_{sym}\, (MeV)$\\
\hline $TM1$&
 $-16.3$&
 $0.658$&
 $0.1455$&
 $281.99$&
 $36.82$\\
\hline $TM1+nonl.$&
 $-16.3$&
 $0.664$&
 $0.1455$&
 $281.99$&
 $32.22$ \\
\hline $GM3$ & $-16.3$ & $0.78$ & $0.153$ & $240$ & $32.5$ \\
\hline
\end{tabular}
\end{center}
\vspace*{0.15in}
\end{table}
 The vector potential for the electromagnetic stress tensor $F_{\mu\nu}$
is given by $A_{\mu }=\{A_{0}=0,\, A_{i}\}$ where
\begin{equation*}
A_{i}=-\frac{1}{2}\varepsilon
_{ilm}x^{l}B_{0}^{m}.
\end{equation*}
The symmetry in which uniform magnetic field ${\textbf B}_{0}$
lies along the z-axis has been chosen chosen
$B_{0}^{m}=(0,0,B_{z})$. Considering the influence of an external
magnetic field on proto-neutron star constituents one can start
with determining the dispersion relations for them. The dispersion
relation for an electron in a magnetic field is
\begin{equation*}
\varepsilon_{e}=\sqrt{p_{z}^{2}+m_{e}^{2}+2neB_{z}},
\end{equation*}
where $n$ is the Landau level, $p_{z}$ is the electron momentum
along the z-axis and $m_{e}$ is the rest mass of the electron
\cite{mil14}. Along the field direction a particle motion is free
and quasi-one-dimensional with the modified density of states
which is given by a sum
\begin{eqnarray*}
2\int \frac{d^{3}p}{(2\pi )^{3}}\rightarrow \sum _{s}\sum
_{n=0}^{\infty }[2-\delta _{n0}]\int \frac{eB_{z}}{(2\pi
)^{2}}dp_{z}, &  &
\end{eqnarray*}
  $\delta _{n0}$ denotes the Kronecker delta \cite{mil14}.
The spin degeneracy equals 1 for the ground ($n=0$) Landau level
and 2 for $n\geq 1$. Similar results can be obtained for muons by
replacing the electron quantities by the corresponding muon
quantities. The energy dispersion relation for protons for
arbitrary Landau level in a magnetic field is \cite{Lati}
\begin{eqnarray}
\varepsilon
_{p,s}=\sqrt{p_{z}^{2}+(\sqrt{M_{p,eff}^{2}+2nQB_{z}}+s\kappa
_{p}B_{z})^{2}} \label{ener}
\end{eqnarray}
and $s=\pm 1$ being "spin" number. For moderate values of
magnetic fields strength this equation
 reduces to the following one
\begin{eqnarray*}
\varepsilon _{p}\sim \sqrt{p_{z}^{2}+M_{p,eff}^{2}+2neB_{z}}.
\end{eqnarray*}
A neutron can interact with an external electromagnetic field by
means of the Pauli non-minimal coupling. Then the energy
dispersion relation for neutrons in a magnetic field is given by
\begin{equation}
\varepsilon _{n,s}=\sqrt{p_{\parallel}^{2}+(\sqrt{M_{n,eff}^{2}+
p_{\perp}^{2}}+s\kappa _{n}B_{z})^{2}}\label{neutron}
\end{equation}
where $p_{\parallel}=p_z$ and $p_{\perp}$ are the components of
the neutron momentum parallel and perpendicular to the magnetic
field.
 As it is usually assumed in quantum
hadrodynamics the mean field approximation is adopted and for the
ground state of homogeneous infinite matter quantum fields
operators are replaced by their classical expectation values. One
can separate mesonic fields into classical mean field values and
quantum fluctuations which are not included in the ground state
and thus mesons fields are replaced by their mean values $\sigma
=<\varphi
>$, $d^{a}=<\delta ^{a}>=\delta ^{a,3} d_{0}$, $w_{\mu }=<\omega _{\mu }>=\delta _{\mu ,0}w_{0}$ and $r_{\mu
}^{a}=<\rho _{\mu }^{a}>=\delta ^{a,3}\delta _{\mu ,0}\, r_{0}$.
The Dirac equation at the mean field level for the nucleon quasi-particle has the form
\begin{equation}
(i\gamma^{\mu}D_{\mu}-M_{N,eff}-\kappa_N\sigma_{\mu\nu}F^{\mu\nu})\psi=0
\end{equation}
with $M_{N,eff}$ being the effective nucleon mass generated by the
nucleon and scalar fields interactions and defined as
\begin{eqnarray}
 & M_{p,eff}=M\delta _{p}=M-g_{s}\sigma -g_{\delta }d_{0}, & \\
 & M_{n,eff}=M\delta _{n}=M-g_{s}\sigma +g_{\delta }d_{0}. &
\end{eqnarray}
The main effect of the inclusion of $\delta$ meson becomes evident
studying properties of proto-neutron star matter especially
nucleon mass splitting and the form of the equation of state. With
the effective nucleon mass $M_{N,eff}$ one can redefine the proton
and neutron chemical potentials
\begin{eqnarray}
 & \mu _{p}=\varepsilon _{p}+g_{\omega }w_0+\frac{1}{2}g_{\rho }r_{0} & \label{mp}\\
 & \mu _{n}=\varepsilon _{n}+g_{\omega }w_0-\frac{1}{2}g_{\rho }r_{0} & \label{mn}
\end{eqnarray}
where $\varepsilon _{p}$ and $\varepsilon _{n}$ is given by
relations (\ref{neutron}) and (\ref{ener}). The obtained effective
nucleon chemical potentials allow to define conditions that are
regarded as essential to uniquely determine the complete
equilibrium composition of proto-neutron star matter at given
density. These conditions arise from charge neutrality, baryon and
lepton number conservation. The last one is strictly connected
with the assumption that proto-neutron star matter is opaque to
neutrinos. Neutrinos are trapped inside the matter and this
through the requirement of $\beta$ equilibrium puts certain limits
on the chemical composition of a proto-neutron star matter. For
the case to be considered the processes of $\beta$-decay and
inverse $\beta$-decay take place \cite{shap}
\begin{equation}
 p+e\, \leftrightarrow \, n+\nu _{e}.  \label{betae}
\end{equation}
For large enough electron Fermi energy it is energetically favorable for electrons to convert to muons
\begin{equation}
\mu +\nu _{e}\, \leftrightarrow \, e+\nu _{\mu }.  \label{betamu}
\end{equation}
The chemical equilibrium established by the above processes is given by
relations between chemical
potentials of proto-neutron star constituents and impose constrains on
them.
\begin{eqnarray}
 & \mu _{\nu _{e}}=\mu _{e}+\mu _{p}-\mu _{n} & \\
 & \mu _{\nu _{\mu }}=\mu _{\nu _{e}}-\mu _{e}+\mu _{\mu }&.
\end{eqnarray}
From the equations (\ref{mp}) and (\ref{mn}) the electron neutrino
chemical potential can be expressed in the following way
\begin{eqnarray}
\label{poten} \mu _{\nu _{e}}=\mu _{e}+\varepsilon
_{p}-\varepsilon _{n}+g_{\rho }r_{0}.
\end{eqnarray}
With the assumption that only electron neutrinos are captured in
the star core $(\mu_{\nu_{\mu}}=0)$ the following relations for electron neutrino and muon
chemical potentials can be obtained
\begin{eqnarray}
\label{relation}
\mu _{\nu _{e}}&=&\mu _{e}+\varepsilon _{p}-\varepsilon _{n}+g_{\rho }r_{0} \nonumber \\
\mu _{\mu }&=&-\mu _{\nu _{e}}+\mu _{e}=-\varepsilon
_{p}+\varepsilon _{n}-g_{\rho }r_{0}.
\end{eqnarray}
Both electron neutrino and muon chemical potentials depend on
nucleon asymmetry. The muon chemical potential depends on the
asymmetry of the whole system through the value of the $\rho$
meson field which is directly connected with the neutron excess
and the difference of nucleon energies. Comparison between the
matter with high asymmetry $(Y_{p}=0.11)$ and the one with low
asymmetry indicates that in the former case the muon chemical
potential is greater than that of the later one. This translates
to the lower number of muons in the proto-neutron star matter with
the proton number $Y_{p}$ of the order of $0.38$. The increase of
neutron excess is connected with the increasing number of muons.
Their contribution to the pressure and energy density of the
system become significant. To construct the equation of state of
the system the energy-momentum tensor has to be calculated. The
pressure is related to the statistical average of the trace of the
spatial components of the energy-momentum tensor
$P=\frac{1}{3}<T_{ii}>$ whereas the energy density $\varepsilon$
equals $<T_{00}>$. In general in the presence of magnetic field
the pressure can be written as a sum of the isotropic part which
includes contributions coming from electrons, nucleons, mesons and
the electromagnetic anisotropic part $T_{\mu \nu }^{B}$ is given
by
\begin{eqnarray*}
T_{\mu \nu }^{B}=F_{\mu }^{\lambda }F_{\nu \lambda
}-\frac{1}{4}g_{\mu \nu }F_{\alpha \beta }F^{\alpha \beta }.
\end{eqnarray*}
In cartesian coordinates the electromagnetic part of the energy
momentum tensor in the flat space-time has the form
\begin{equation}
T_{\mu \nu }^{B}=\left(\begin{array}{cccc}
 \frac{1}{2}B^{2} & 0 & 0 & 0\\
 0 & \frac{1}{2}B^{2} & 0 & 0\\
 0 & 0 & \frac{1}{2}B^{2} & 0\\
 0 & 0 & 0 & -\frac{1}{2}B^{2}\end{array}
\right).\label{tensor metryczny}
\end{equation}
 On the other hand in polar coordinates $T_{\mu \nu }^{B}$ has been written as
follows
\begin{eqnarray*}
\left(\begin{array}{cccc}
 \frac{1}{2}B^{2} & 0 & 0 & 0\\
 0 & -\frac{1}{2}B^{2}\cos 2\theta  & \frac{1}{2}B^{2}r\sin 2\theta  & 0\\
 0 & \frac{1}{2}B^{2}r\sin 2\theta  & \frac{1}{2}B^{2}r^{2}\cos 2\theta  & 0\\
 0 & 0 & 0 & \frac{1}{2}B^{2}r^{2}\sin ^{2}\theta \end{array}
\right). &  &
\end{eqnarray*}
Taking the average over all directions allows to obtain the
isotropic form of the energy momentum tensor relevant for the
spherical symmetric configuration
\begin{eqnarray*}
\left(\begin{array}{cccc}
 \frac{1}{2}B^{2} & 0 & 0 & 0\\
 0 & \frac{1}{6}B^{2} & 0 & 0\\
 0 & 0 & \frac{1}{6}B^{2} & 0\\
 0 & 0 & 0 & \frac{1}{6}B^{2}
 \end{array}
\right). &  &
\end{eqnarray*}
The assumption of the positivity of the total pressure $P$ has
been made. Any negative contributions to the pressure diminish it
and lead to the decrease of a star radius. The described above
averaging
 approximates the star as an spherically
symmetric object with the radius $R$ which is smaller than the
equatorial radius of an anisotropic star. (The case of an
anisotropic star is calculated in details in the paper by Konno et
al. \cite{konno}). The condition of the positivity of the total
pressure establishes a limiting value of the density noted as
$\rho _{crit}$ and at the same time puts limit on the spherical
symmetric approximation (the deformation of the star is
neglected). For densities smaller than the critical one the total
pressure became negative and there is no gravity compensation in
this direction. The total pressure and the energy density of the
system supplemented by corrections coming from magnetic field are
obtained by adding contributions from mesons and fermions
\begin{eqnarray*}
& P=P_{f}+P_{QED}-U(\sigma)+\frac{1}{2}M_{\rho}^{2}r_{0}^{2}
-\frac{1}{2}M_{\omega}^{2}w_0^{2
}-\frac{1}{2}M_{\delta}^{2}d_0^{2}+\Lambda_{v}g_{\omega}^{2}g_{\rho}^{2}r_{0}^{2}w_{0}^{2
}+ & \\
& +
g_{\sigma}^{2}g_{\rho}^{2}\Lambda_{4}r_{0}^{2}\sigma^{2}+\frac{1}{4}c_{3}w_{0}^{4}+
\frac{1}{24}\zeta g_{\rho}^{4}r_{0}^{4} & \\
& \varepsilon  =
\varepsilon_{f}+\varepsilon_{QED}+U(\sigma)+\frac{3}{4}c_{3}w_{0}^{4}
+\frac{1}{2}M_{\omega}^{2}w_{0}^{2}+\frac{1}{2}M_{\rho}^{2}r_{0}^{2}+3\Lambda_{v}
g_{\omega}^{2}g_{\rho}^{2}r_{0}^{2}w_{0}^{2}+
g_{\sigma}^{2}g_{\rho}^{2}& \\ & \Lambda_{4}r_{0}^{2}\sigma^{2}+
\frac{1}{8}\zeta g_{\rho}^{4}r_{0}^{4} + \frac{1}{2}M_{\delta}^{2}d_0^{2} & \\
\end{eqnarray*}
where $f=(i,n),\,\,i=(e,\mu,p)$ and $\varepsilon
 _{QED}=\frac{1}{2}B^{2},P_{QED}=\frac{1}{6}B^{2}$.
The fermion pressure is defined as
\begin{eqnarray*}
 & P_{i}=\frac{\gamma _{e}m_{i}^{4}}{4\pi ^{2}}\sum _{s}\sum _{n=0}^{\infty }[2-\delta _{n0}]\{I_{-1,2,+}(z_{i}/t_{i},\varsigma _{i}^{2}+2\gamma _{e}n) & \\
 & +I_{-1,2,-}(z_{i}/t_{i},\varsigma _{i}^{2}+2\gamma _{e}n)\} &
\end{eqnarray*}
 where the Fermi integral
 \begin{equation}
I_{\lambda ,\eta \pm }(u,\alpha )=\int \frac{(\alpha +x^{2})^{\lambda /2}x^{\eta }dx}{e^{\left(\sqrt{\alpha +x^{2}}\mp u\right)}+1}
\end{equation}
was used, $z_{i}=\mu _{0}/m_{i}$, $t_{i}=k_{B}T_{0}/m_{i}$,
$u_{i}=z_{i}/t_{i}$, $\gamma _{e}=B_{z}/B_{c}^{e}$ and
$\varsigma_{i}=m_{i,eff}/m_{i}$. The last quantity equals 1 for
electrons and muons. The critical magnetic field strength for
electron equals $B_{c}^{e}$=$m_{e}^{2}/\mid e\mid $= $4.414\times
10^{13}$ $G$. The fermion energy density is also defined with the
use of the Fermi integral
\begin{eqnarray*}
 & \varepsilon_{i} =\frac{\gamma _{e}m_{i}^{4}}{4\pi ^{2}}\sum _{s}\sum _{n=0}^{\infty }[2-\delta _{n0}]
 \{I_{1,0,+}(z_{i}/t_{i},\varsigma _{i}^{2}+2\gamma _{e}n) & \\
 & +I_{1,0,-}(z_{i}/t_{i},\varsigma _{i}^{2}+2\gamma _{e}n)\}. &
\end{eqnarray*}
\begin{figure}
\includegraphics[width=12cm]{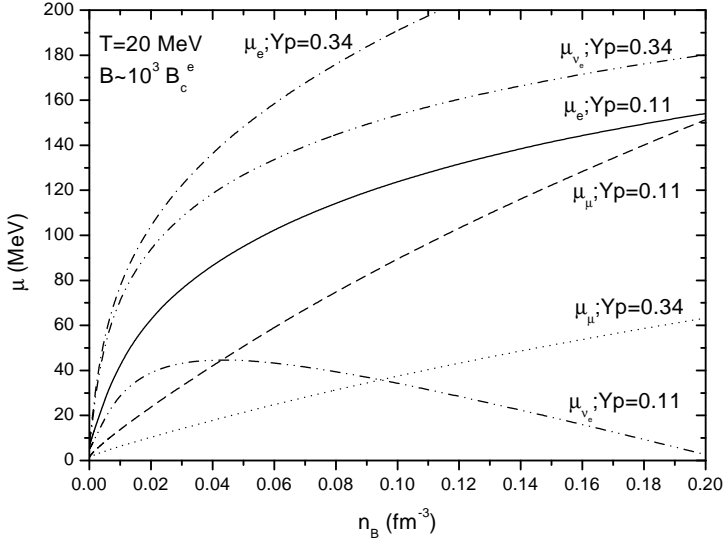}
\label{rys1} \caption{The lepton chemical potentials ($e$, $\mu$,
$\nu_{e} $) as a function of the baryon density.}
\end{figure}
\begin{figure}
\includegraphics[  width=12cm]{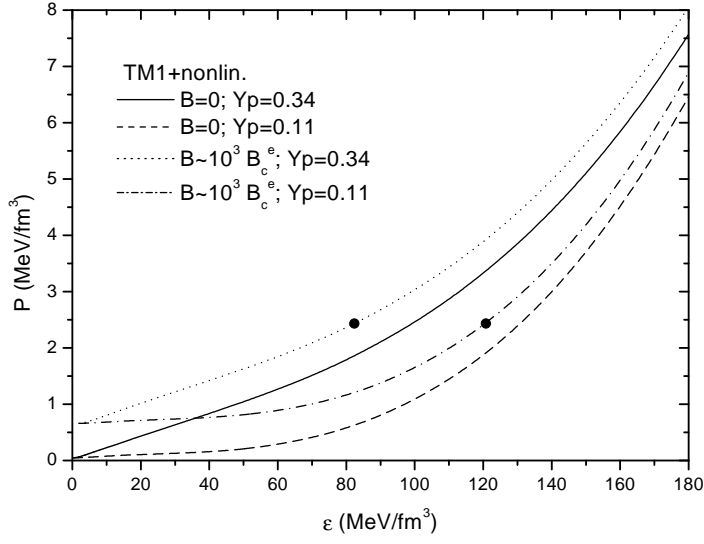}
\label{rys2} \caption{The equation of state for the proto-neutron
star for different values of the proton concentration $Y_{p}$ and
strength of magnetic field. TM1 model with the nonlinear terms.}
\end{figure}
\begin{figure}
\includegraphics[  width=12cm]{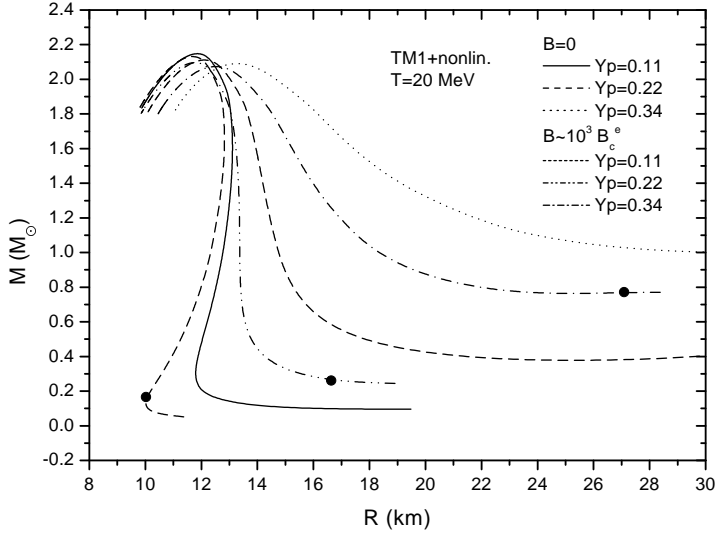}
\label{rys3} \caption{The mass radius relations for different
proton fractions $Y_{p}$ ($T=20$, $MeV$,  $B\sim 10^{3}$
$B_{c}^{e}$, $B_{c}^{e}$=$4.414\times 10^{13}$). For comparison
the mass radius relation for the proto-neutron star without
magnetic field ($B=0$) is also presented).}
\end{figure}
\begin{figure}
\includegraphics[  width=12cm]{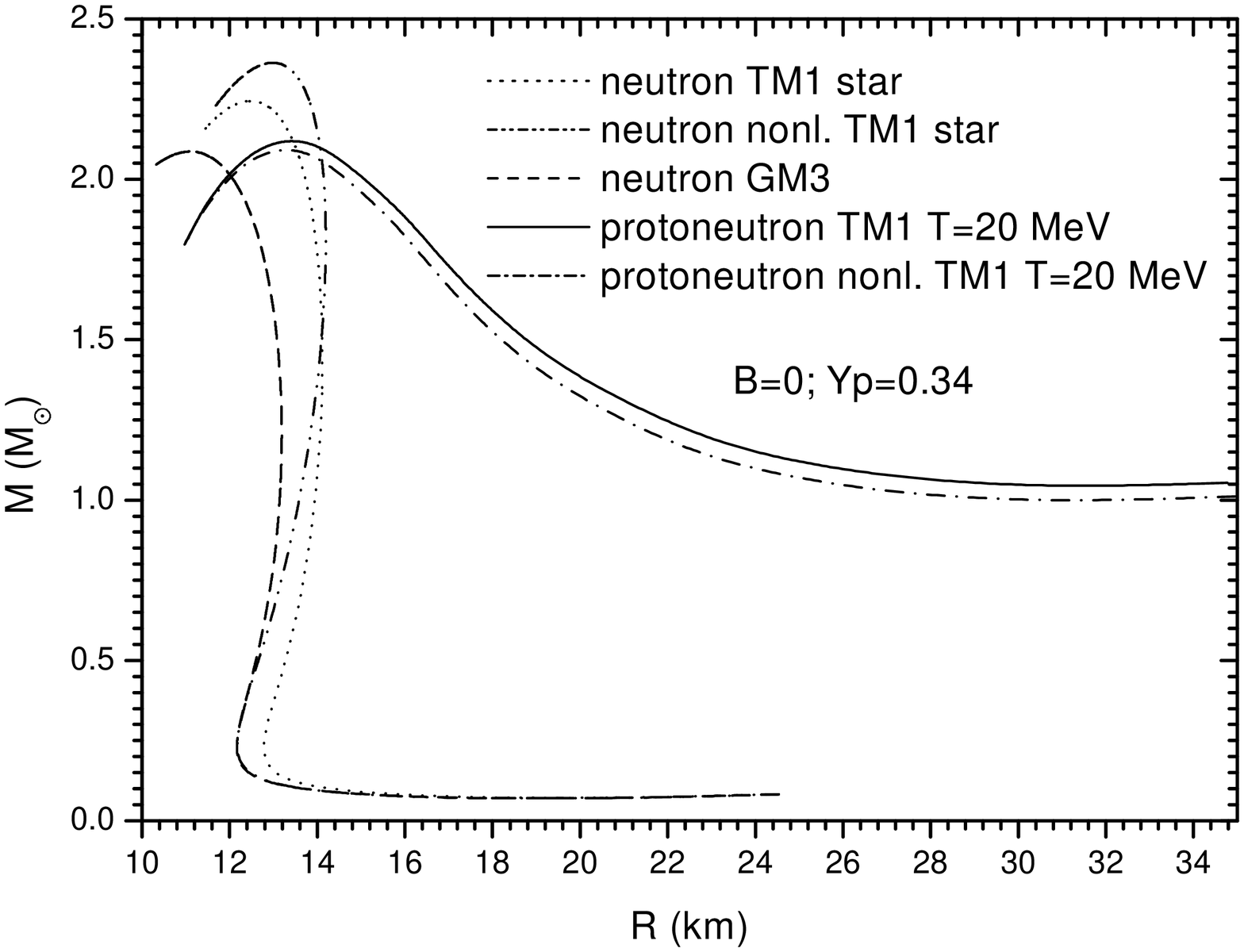}
\label{rys4 } \caption{The mass radius relations for neutron and
proto-neutron stars for different parameter sets.}
\end{figure}
\section*{Conclusion}
Analyzing properties of proto-neutron star matter some features
which are relevant for proto-neutron stars have to be considered.
It is a necessary assumption that the matter is opaque to
neutrinos. As a result neutrinos are trapped dynamically inside
the matter on the diffusion time scale and in contradiction to
neutron star matter the total lepton number $Y_{L}=Y_{e}+Y_{\mu
_{e}}+Y_{\mu }$ is kept constant. Theoretical predictions
determine the value of $Y_{L}$ at the onset of trapping at about
0.35-0.4. Since a proto-neutron star is formed during a supernova
explosion it is safely to make an assumption  about finite
temperature in the interior of a star. This is limited by the
requirement of constant value of entropy in a proto-neutron star
core. In this paper the temperature is fixed and equals 20 MeV.
Thus conditions which have to be considered namely: finite
temperature, lepton number conservation, $\beta$- equilibrium,
charge neutrality and additionally nonzero magnetic field allow to
construct a  realistic proto-neutron star model. All of these
mentioned above conditions influence medium properties such as
nucleon effective masses, density and proton-neutron asymmetry and
in consequence constrain the nucleon and lepton species inside the
star core. The proton fraction $Y_p=n_p/(n_p+n_n)$ which is
determined by $\beta$ equilibrium, for matter transparent to
neutrinos (neutron star matter) takes the value of the order of
$\sim 0.1$ whereas when neutrinos do contribute they displace the
equilibrium proton fraction to the higher value $Y_{p} \sim
0.3-0.36$. For the purpose of comparison all calculations have
been performed for decreasing value of proton fractions
$Y_{p}=0.34,Y_{p}=0.22, Y_{p}=0.11$, respectively. This changing
proton-neutron asymmetry through relation (\ref{relation}) alters
the lepton chemical potentials. Figure 1 displays chemical
potentials of electron neutrinos, electrons and muons for two
extreme values of $Y_{p}$ ($Y_{p}=0.11$ and $Y_{p}=0.34$). They
have been examined for moderate strength of magnetic field. If the
proton fraction equals $Y_{p}=0.34$ (the proto-neutron star
matter) electron and neutrino electron chemical potentials are
increasing functions of a density. Neutrinos are trapped inside
the matter and their chemical potential differs from zero (see
relation \ref{poten}). For the proton fraction $Y_{p}=0.11$ the
matter inside the star resembles that of a neutron star which is
characterized by larger value of asymmetry. Neutrinos are about to
leave the system and their chemical potential vanishes
$\mu_{\nu}\rightarrow 0$. This picture shows the profile of
neutrino chemical potential in the core of the star for the chosen
values of $Y_{p}$ and for chosen parameter set (TM1+nonl.). For
the star with the value of asymmetry typical for a proto-neutron
star matter $(Y_{p} \sim 0.38)$ muons are of insignificant
meaning. However, they become much more important for lower value
of proton number. The asymmetry entered into calculations at
different levels starting from the displacement of proton and
neutron chemical potentials $\mu_{p}\neq \mu_{n}$. In a model with
$\delta$ meson exchange additionally $M_{p}\neq M_{n}$. There are
also contributions coming from nonlinear vector meson
interactions. The nuclear asymmetry diminishes the neutrino
chemical potential and thus lowers neutrino density. The influence
of the asymmetry is most distinctive in the high density region
inside a proto-neutron star. The forms of neutrino profiles in the
proto-neutron star core depend on the asymmetry of the whole
system. The behavior of chemical potentials translates  to the
forms of the equation of state and successively to the mass-radius
relations. Results are obtained for two particular cases namely
for the nonmagnetic one ($B=0$) and for the value of magnetic
field $B \sim 10^{3}B_{c}^{e}$. The presence of magnetic field
makes the equation of state stiffer (Fig. 2), whereas for more
asymmetric matter the obtained equation of state is softer. Dots
represent the limit of making the approximation of spherically
symmetric star in the presence of magnetic field. For more
asymmetric matter the dot is shifted towards higher densities. In
Fig. 3 one can compare the influence of changing proton fraction
on star parameters. The obtained sequences of mass-radius for
nonmagnetic and magnetic proto-neutron stars show that the
increasing asymmetry diminishes proto-neutron star radii causing
that the star parameters start to resemble that of neutron stars.
This effect is even stronger in the presence of magnetic field.
There are relatively minor changes in the value of maximal masses
caused by the variation of proton fraction. However, there is a
tendency of the increase of the maximal mass for growing
asymmetry. The star parameters are also sensitive to the presence
of nonlinear vector meson interaction terms in the Lagrangian
function (\ref{lagrangian}). This is clearly visible in Fig. 4.
Comparison of results obtained for original GM3 and TM1 equations
of state (for neutron stars) with that obtained for TM1+nonl.
shows that the last one gives the configuration with the bigger
value of maximal mass. In the range of lower densities the
TM1+nonl. parameter set gives the star with the smaller radius.
Thus in the case of high asymmetry nonlinearities create more
compact objects. Nonlinearities alter proto-neutron star
parameters insignificantly. This is caused by lower value of the
asymmetry.

\end{document}